\documentclass[epj,final]{svjour}
\usepackage{graphicx}
\usepackage[usenames]{color}
\usepackage{amssymb}
\usepackage{amsmath}
\DeclareMathAccent{\pol}{\mathord}{letters}{"7E}

\begin{document}
\definecolor{MyHQM}{rgb}{0.57,0,1}
\definecolor{MyCapstick}{rgb}{0,0.7,0.7}
%
\title{Lowest $Q^2$ Measurement of the $\gamma^*p\rightarrow\Delta$ Reaction: Probing the Pionic Contribution}

\newcommand{\mitlnsadd}{Department of Physics, Laboratory
for Nuclear Science and Bates Linear Accelerator Center, Massachusetts Institute of Technology, Cambridge,
Massachusetts 02139, USA}

\newcommand{\mainzadd}{Institut f\"ur Kernphysik, Johannes Gutenberg-Universit\"at Mainz, D-55099  Mainz, Germany}

\newcommand{\rikenadd}{Current address: Radiation Laboratory, RIKEN, 2-1 Hirosawa, Wako, Saitama 351-0198, Japan}

\newcommand{\bonnadd} {Current address: Univ. Bonn, Physikalisches Inst.,
Nussallee 12, D-53115 Bonn, Germany}

\newcommand{\athensadd}{Institute of Accelerating Systems and Applications and
  Department of Physics, University of Athens, Athens, Greece}

\newcommand{\zagrebadd}{Department of Physics, University of Zagreb, Croatia}

\newcommand{\ukadd}{Department of Physics and Astronomy, University of
  Kentucky, Lexington, Kentucky 40206 USA}

\newcommand{\uljadd}{Institute Jo\v zef Stefan, University of Ljubljana,
  Ljubljana, Slovenia}

\newcommand{\dukeadd}{Current address: Department of Physics, Duke
  University/TUNL, Durham, North Carolina 27708, USA}

\newcommand{\mitlns}{1}
\newcommand{\mainz}{2}
\newcommand{\uk}{3}
\newcommand{\athens}{4}
\newcommand{\zagreb}{5}
\newcommand{\ulj}{6}

\author{The A1 Collaboration \\\\
S. Stave\inst{\mitlns}$^{,}$\thanks{\dukeadd} \and
M.~O.~Distler\inst{\mainz} \and
I.~Nakagawa\inst{\mitlns}$^{,}$\inst{\uk}$^{,}$\thanks{\rikenadd} \and
N. Sparveris\inst{\athens} \and
P.~Achenbach\inst{\mainz} \and
C.~Ayerbe Gayoso\inst{\mainz} \and
D.~Baumann\inst{\mainz} \and
J.~Bernauer\inst{\mainz} \and
A. M. Bernstein\inst{\mitlns} \and
R.~B{\"o}hm\inst{\mainz} \and
D.~Bosnar\inst{\zagreb} \and
T. Botto\inst{\mitlns} \and
A.~Christopoulou\inst{\athens} \and
D.~Dale\inst{\uk} \and
M.~Ding\inst{\mainz} \and
L.~Doria\inst{\mainz} \and
J.~Friedrich\inst{\mainz} \and
A.~Karabarbounis\inst{\athens} \and
M.~Makek\inst{\zagreb} \and
H.~Merkel\inst{\mainz} \and
U.~M{\"u}ller\inst{\mainz} \and
R.~Neuhausen\inst{\mainz} \and
L.~Nungesser\inst{\mainz} \and
C.N.~Papanicolas\inst{\athens} \and
A.~Piegsa\inst{\mainz} \and
J.~Pochodzalla\inst{\mainz} \and
M.~Potokar\inst{\ulj} \and
M.~Seimetz\inst{\mainz}$^{,}$\thanks{\bonnadd} \and
S.~{\v S}irca\inst{\ulj} \and
S.~Stiliaris\inst{\athens} \and
Th.~Walcher\inst{\mainz} \and
M.~Weis\inst{\mainz}
}

%
%
\institute{
\mitlnsadd \and
\mainzadd \and
\ukadd \and
\athensadd \and
\zagrebadd \and
\uljadd
}

\mail{A.~M.~Bernstein, e-mail: bernstein@lns.mit.edu}

\authorrunning{S. Stave \emph{et al.}}
\titlerunning{Lowest $Q^2$ Measurement of the $\gamma^*p\rightarrow\Delta$ Reaction: Probing the Pionic Contribution} 

\date{Received: date / Revised version: date}
%
\abstract{
To determine nonspherical angular momentum amplitudes in hadrons at long ranges (low $Q^{2}$), data were taken 
for the $p(\pol{e},e'p)\pi^0$ reaction in the $\Delta$ region at
$Q^2=0.060$ (GeV/c)$^2$ utilizing the magnetic spectrometers of the A1
Collaboration at MAMI.  The results for the dominant transition
magnetic dipole amplitude  and the quadrupole to dipole ratios  at
$W=1232$ MeV are:
$M_{1+}^{3/2}=(40.33 \pm 0.63_{\rm stat+syst} \pm 0.61_{\rm
  model})(10^{-3}/m_{\pi^+})$,
Re($E_{1+}^{3/2}/M_{1+}^{3/2})=(-2.28 \pm 0.29_{\rm
  stat+syst} \pm 0.20_{\rm model}$)\%, and 
Re($S_{1+}^{3/2}/M_{1+}^{3/2})=(-4.81 \pm 0.27_{\rm
  stat+syst} \pm 0.26_{\rm model}$)\%.  
These disagree with predictions of constituent  quark models but are in reasonable agreement with lattice calculations with non-linear (chiral) pion mass extrapolations, with chiral effective field  theory, and with
dynamical models with pion cloud effects. These results confirm the dominance, and general $Q^{2}$ variation, of the  pionic contribution at large distances.
}
\maketitle

Experimental confirmation of the presence of nonspherical hadron amplitudes (i.e. d states in quark models or p wave $\pi$-N states) is fundamental and has been the subject of intense experimental and theoretical interest (for reviews see \cite{nstar2001,cnp,amb}). This effort has focused on the measurement of the electric and Coulomb quadrupole amplitudes (E2, C2) in the predominantly M1 (magnetic dipole-quark spin flip) $\gamma^* N
\rightarrow \Delta$ transition. Measurements of the E2 amplitude from photopion reaction experiments with polarized photons
have been reported \cite{beck,blanpied}. Electroproduction experiments at JLab \cite{joo,frolov} for $Q^2$ from 0.4 to 4.0 (GeV/c)$^2$, at Bates \cite{warren,mertz,kunz,sparveris} at $Q^2 = 0.127$
(GeV/c)$^2$  and Mainz \cite{pospischil,bartsch,elsner} at $Q^2 = 0.127, 0.20$ (GeV/c)$^2$ have been published.
The present result at $Q^2 = 0.060$ (GeV/c)$^2$ is the lowest
$Q^{2}$ value probed to date in modern, precision electroproduction. It adds a very important point to determine the
physical basis of long range nucleon and $\Delta$ nonspherical amplitudes and is a test of the $Q^{2}$ region where pionic effects are predicted to be dominant and appreciably changing.

Since the proton has spin 1/2, no quadrupole moment can be
measured.  However, the $\Delta$ has spin 3/2 so the 
$\gamma^*N\rightarrow \Delta$ reaction can be studied for quadrupole amplitudes in the nucleon and $\Delta$.  
Due to spin and parity conservation in the $\gamma^*N(J^\pi=1/2^+)
\rightarrow \Delta(J^\pi=3/2^+)$ reaction, only three multipoles can
contribute to the transition: the magnetic dipole (M1), the electric
quadrupole (E2), and the Coulomb quadrupole (C2) photon absorption
multipoles.  The corresponding resonant pion production multipoles are 
$M_{1+}^{3/2}$, $E_{1+}^{3/2}$, and $S_{1+}^{3/2}$.  The relative quadrupole to dipole ratios are
${\rm EMR}={\rm Re}(E_{1+}^{3/2}/M_{1+}^{3/2})$ and 
${\rm CMR}={\rm Re}(S_{1+}^{3/2}/M_{1+}^{3/2})$.
In the quark model, the nonspherical amplitudes in the 
nucleon and $\Delta$ are caused by the non-central, tensor interaction
between quarks \cite{glashow}.
However, the magnitudes of this effect for the predicted E2 and C2  amplitudes\cite{capstick_karl} are at least an
order of magnitude too small to explain the experimental results (see Fig.~\ref{fig:Q06} below) and
even the dominant M1 matrix element is $\simeq$ 30\% low
\cite{amb,capstick_karl}. A likely cause of these dynamical shortcomings is that the quark model
does not respect chiral symmetry, whose spontaneous breaking leads to strong emission of virtual pions (Nambu-Goldstone Bosons)\cite{amb}. These couple
to  nucleons as $\pol{\sigma}\cdot \pol{p}$ where  $\pol{\sigma}$ is the nucleon spin, and $\pol{p}$ is the pion momentum. The coupling is strong in the p wave and mixes in non-zero
angular momentum components. Based on this,
it is physically reasonable to expect that the pionic contributions increase the M1  and 
dominate the E2 and C2 transition matrix elements in the low $Q^2$
(large distance) domain. This was first indicated by adding pionic
effects to quark models\cite{quark_pion},
 subsequently shown in pion cloud model calculations\cite{sato_lee,dmt}, and 
recently demonstrated in chiral effective field theory calculations\cite{gail_hemmert,pasc}.

The five-fold differential cross section for the
$p(\pol{e},e'p)\pi^{0}$ reaction is written as five two-fold
differential cross sections with an explicit $\phi^*$
dependence as \cite{drechsel_tiator}

\begin{eqnarray}
\frac{d^5\sigma}{d\Omega_f dE_f d\Omega} & = & \Gamma (\sigma_T + \epsilon
\sigma_L + v_{LT}\sigma_{LT} \cos\phi^* \nonumber \\ 
& + & \epsilon \sigma_{TT} \cos 2\phi^* 
 + h p_{e} v_{LT'}\sigma_{LT'} \sin \phi^*)
\end{eqnarray}
where $\epsilon$ is the transverse polarization of the virtual photon,
$v_{LT}=\sqrt{2\epsilon(1+\epsilon)}$, $v_{LT'}=\sqrt{2\epsilon(1-\epsilon)}$,
$\Gamma$ is the virtual photon flux, $\phi^*$  is the pion center of mass
azimuthal angle with respect to the electron scattering plane, $h$ is the electron helicity, and $p_{e}$ is the magnitude of the electron longitudinal polarization. 
The virtual photon differential cross sections
($\sigma_{T},\sigma_{L},\sigma_{LT},\sigma_{TT},\sigma_{LT'}$) 
are all functions of the
center of mass energy $W$, the four momentum transfer squared $Q^2$,
and the pion center of mass polar angle $\theta_{\pi q}^{*}$ (measured
from the momentum transfer direction). They are bilinear combinations of the multipoles \cite{drechsel_tiator}.

The $p(\pol{e},e'p)\pi^{0}$ measurements were performed using the A1 spectrometers at the Mainz Microtron\cite{blom}.  Electrons were
detected in Spectrometer A  and  protons in  Spectrometer B.  
Timing and missing mass cuts were
sufficient to eliminate the $\pi^-$ background.  
Spectrometer B has the ability to measure at up
to 10 degrees out-of-plane in the lab. Due to the Lorentz boost this is significantly larger  in the center of mass frame.  
The Mainz Microtron delivers a longitudinally polarized, continuous, 855 MeV beam. Beam polarization
was measured periodically with a M\o{}ller polarimeter to be $\approx 75\%$. The beam of up to 25 $\mu$A was
scattered from a liquid hydrogen cryogenic target. 
Sequential measurements were
made at $W=1221$ MeV, $Q^2= 0.060$ GeV$^2$/c$^2$, and
$\theta_{\pi q}^{*}=143,156,180^\circ$.  For non-parallel measurements
($\theta_{\pi q}^{*}\ne 180$), the proton arm was moved through
three $\phi^*$ settings while keeping the $\theta_{\pi q}^*$ value
constant. The spectrometers were aligned in the lab with a precision of
0.6 mm and 0.1 mrad 
with a central momentum resolution of 0.01\% and
angular resolution at the target of 3 mrad \cite{blom}.  
The beam energy has an absolute uncertainty of $\pm 160$ keV and a spread of 30 keV (FWHM)
\cite{blom}.  The effects of these uncertainties and the various kinematic cuts were studied  to estimate an overall
systematic error for the cross sections  of 3 to 4\%. 
This was tested with elastic electron-proton scattering and the data agree with a fit to the world data \cite{mergell} at the 3\% level.

With measurements at three $\phi_{\pi q}^*$ values at a fixed
$\theta_{\pi q}^{*}$ and using a polarized electron  beam, the  cross sections  $\sigma_0=\sigma_T +
\epsilon\sigma_L, \sigma_{TT}, \sigma_{LT}$, and $\sigma_{LT'}$ can be extracted from the $\phi^*$ and beam helicity dependence of the cross section. 
Care was taken to ensure good
kinematic overlap between the different angular settings.  Typically the phase space overlaps were $\Delta W \approx40$ MeV, $\Delta Q^2\approx0.04$ GeV$^2$/c$^2$, $\Delta
\theta_{\pi q}^{*}\approx10^\circ$, and $\Delta
\phi_{\pi q}^{*}\approx40^\circ$.  
Since the cross
sections vary across the spectrometer acceptance, 
the shape of the cross section
given by several models was used to refer all of the points to the
center of the acceptance.  This is a small correction (typically 3\%)
and depends 
only on the relative cross sections across the spectrometer acceptance. Several models were used for the
collapse and each gives results consistent at the 0.5\% level \cite{stave_thesis}. 

\begin{figure*}
\begin{center}
\includegraphics[angle=0,height=6cm,width=16cm]{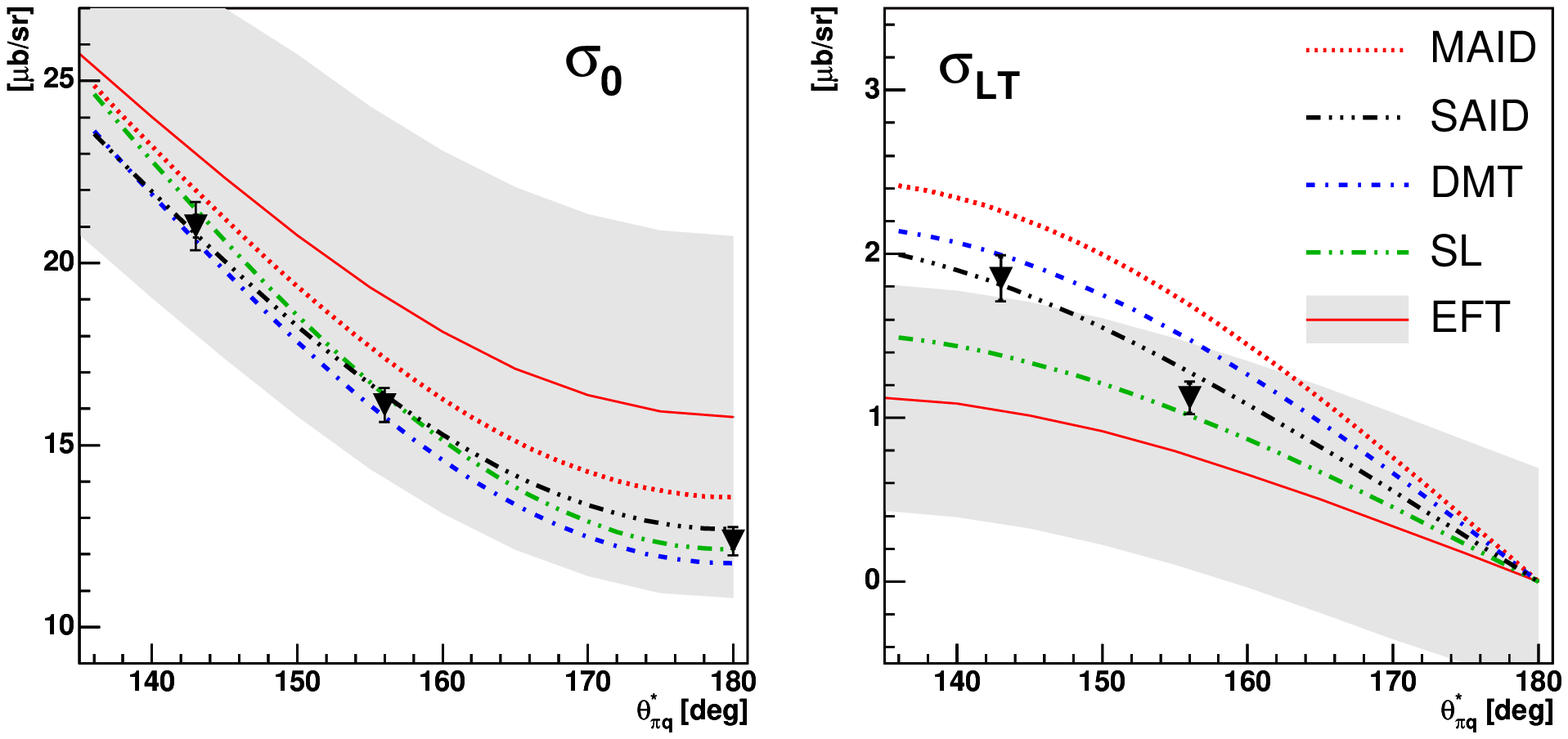}
\hrule
\includegraphics[angle=0,height=6cm,width=16cm]{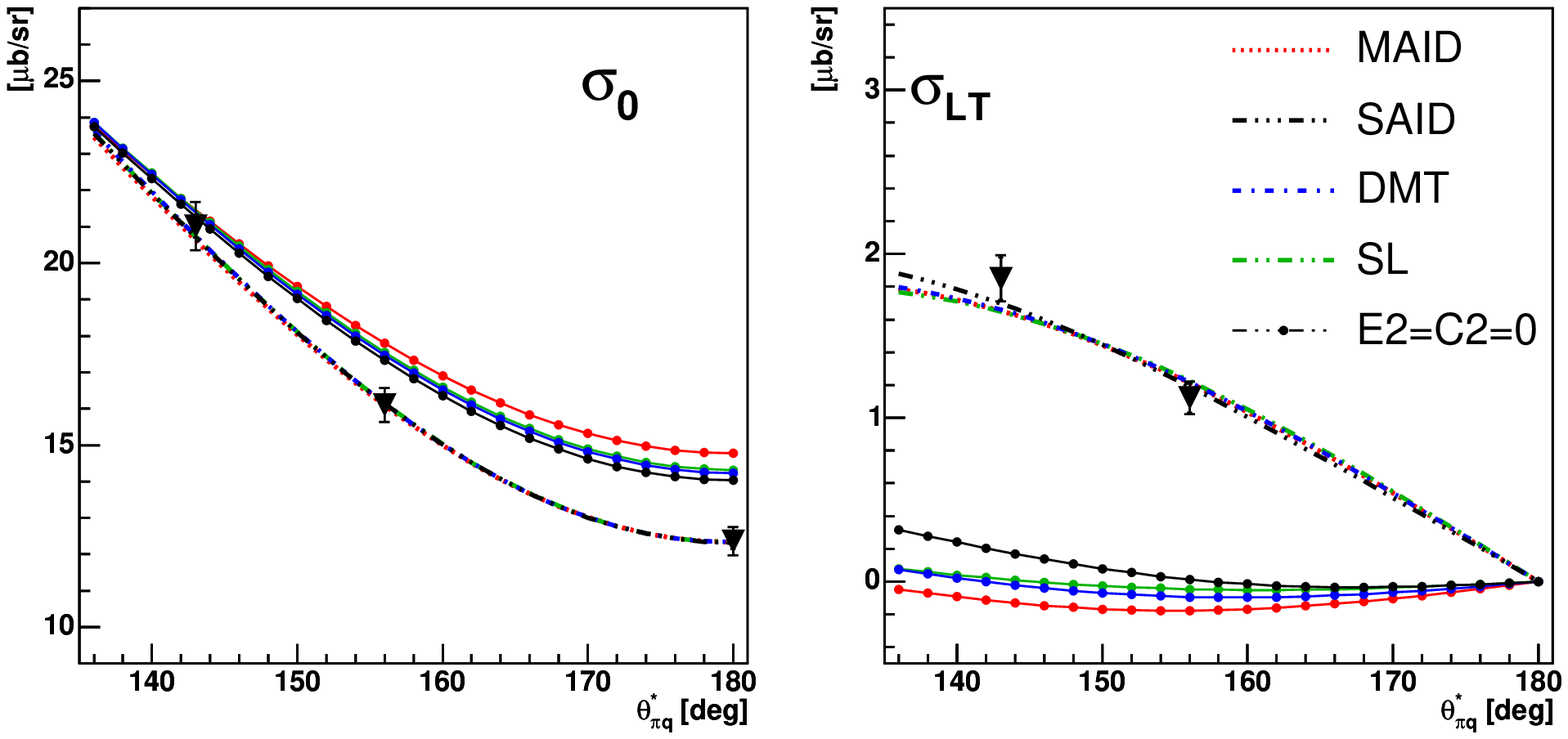}
\end{center}
\caption{\label{fig:XS:Sig0_LT}(Color online) The measured $\sigma_0=\sigma_T +
\epsilon\sigma_L$ and $\sigma_{LT}$
differential cross sections as a function of $\theta_{\pi q}^{*}$ at
$W=1221$ MeV and $Q^2=0.060$ (GeV/c)$^2$ before (top panels) and after (bottom panels) fitting. 
 The $\blacktriangledown$ symbols are our data
points and include the experimental and model errors (see
Table~\ref{table1}) 
added in quadrature. The EFT predictions  \cite{pasc}
are plotted with their estimated uncertainties. The other curves
represent predictions from the  MAID 2003 \cite{maid1},
SL(Sato-Lee)\cite{sato_lee}, DMT \cite{dmt}, and SAID \cite{said}
models.  The lines with dots are the fitted models with the $E_{1+}^{3/2}$ and $S_{1+}^{3/2}$ quadrupole terms set to zero.} 
\end{figure*}

\begin{figure*}
\begin{center}
\includegraphics[angle=0,height=6cm,width=16cm]{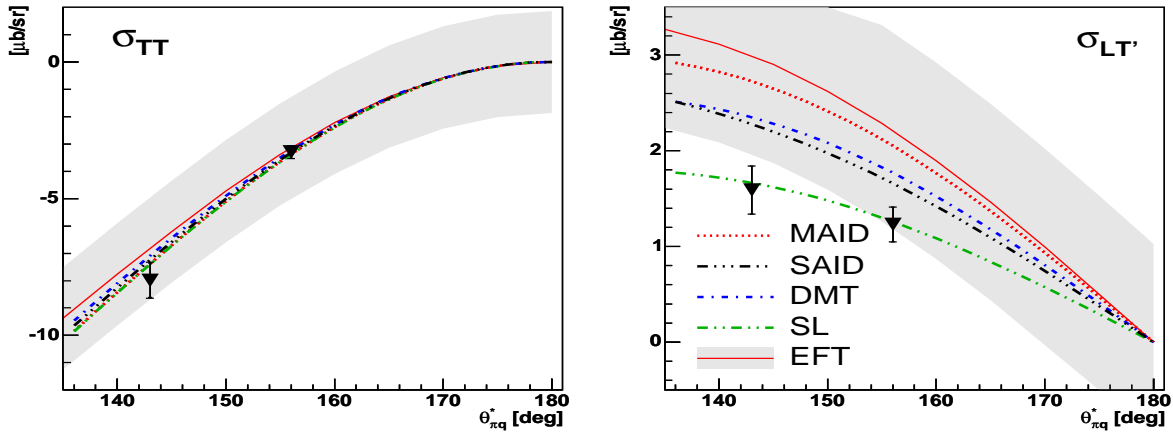}
\end{center}
\caption{\label{fig:XS:SigTT_LTp}(Color online) The measured $\sigma_{TT}$ and $\sigma_{LT'}$
differential cross sections as a function of $\theta_{\pi q}^{*}$ at
$W=1221$ MeV and $Q^2=0.060$ (GeV/c)$^2$.
 The $\blacktriangledown$ symbols are our data
points and include the experimental and model errors (see
Table~\ref{table1}) 
added in quadrature. The EFT predictions  \cite{pasc}
are plotted with their estimated uncertainties. The other curves
represent predictions from the  MAID 2003 \cite{maid1},
SL(Sato-Lee)\cite{sato_lee}, DMT \cite{dmt}, and SAID \cite{said}
models.  The model curves after fitting are almost identical to those
before and so
have been suppressed.  See text for details. } 
\end{figure*}

The measured partial cross sections are plotted  in 
Figs.~\ref{fig:XS:Sig0_LT} and \ref{fig:XS:SigTT_LTp}. Figure~\ref{fig:XS:Sig0_LT}
shows $\sigma_{0}$ and $\sigma_{LT}$ with the chiral effective field
theory (EFT) predictions \cite{pasc} which have uncertainties which
reflect an estimate of the neglected higher order terms in the
chiral expansion. The other models are the phenomenological model MAID
2003\cite{maid1}, the pion cloud dynamical models of Sato and
Lee\cite{sato_lee} and of DMT\cite{dmt} (Dubna, Mainz, and Taipei),
and the SAID multipole analysis\cite{said}. There is a significant
spread in these model calculations due to differences in the resonant
and background amplitudes. However it is impressive  that the four
model curves almost fall on top of each other  when  the three
resonant $\gamma^{*} p \rightarrow \Delta$ amplitudes
($M_{1+}^{3/2},E_{1+}^{3/2},S_{1+}^{3/2}$) are varied to fit the data
as shown in the lower panel of Fig.~\ref{fig:XS:Sig0_LT}. 
In addition, this panel shows
the ``spherical'' calculated curves when the  resonant quadrupole
amplitudes ($E_{1+}^{3/2}$ in $\sigma_{0}$ and $S_{1+}^{3/2}$ in
$\sigma_{LT}$) are set equal to zero. The difference between the
spherical and full curves shows the sensitivity of these cross sections
to the quadrupole amplitudes and demonstrates the basis of the
present measurement. The small spread in the spherical curves
indicates their sensitivity to the model dependence of the background
amplitudes.

Figure~\ref{fig:XS:SigTT_LTp} shows the measured cross sections for $\sigma_{TT}$ and
$\sigma_{LT'}$ with the same model curves as in Fig.~\ref{fig:XS:Sig0_LT}.
 It is
seen that within the relatively large estimated uncertainties the
chiral effective field theory calculations\cite{pasc} are consistent
with experiment. For $\sigma_{TT}$ the four model
calculations\cite{maid1,sato_lee,dmt,said} are in good agreement
with each other and with the data. However for $\sigma_{LT'}$ this is
not the case and only the Sato-Lee model agrees with experiment. 

It is instructive to examine why  $\sigma_{LT}$ is sensitive to the Coulomb quadrupole amplitude,  ${\rm Im}[S_{1+}]$, and   $\sigma_{LT'}$ is primarily sensitive to the background. The time reversal even observable $\sigma_{LT}$  contains the
interference amplitude ${\rm Re}[S_{1+}^* M_{1+}]$ 
which is primarily sensitive to ${\rm Im}[S_{1+}] {\rm Im}[M_{1+}] $ 
where the latter is the dominant multipole amplitude. 
By contrast the time reversal odd observable $\sigma_{LT'}$ contains
${\rm Im} [S_{1+}^* M_{1+}]$. This is primarily sensitive to 
${\rm Re}[S_{1+}] {\rm Im}[M_{1+}]$  and
therefore does not measure the Coulomb quadrupole amplitude but
rather is sensitive to a background term times the dominant magnetic
dipole term. The details will be presented in a future publication\cite{stave_etal}.

On the other hand, the other
cross sections show  a minimum of model dependence  and can be used to
accurately extract the three resonant amplitudes as will be discussed
below.
The model curves after fitting for $\sigma_{TT}$ and $\sigma_{LT'}$
are almost identical to those before and so
have been suppressed.
$\sigma_{LT'}$ is insensitive to the resonant parameters, as mentioned
above, and
$\sigma_{TT}$ while sensitive to $E_{1+}$, is dominated by the
$M_{1+}$ term.  

As has been discussed above, we have obtained the values of the three
resonance amplitudes ($M_{1+}^{3/2},E_{1+}^{3/2},S_{1+}^{3/2}$) using
fits with four reaction models\cite{maid1,sato_lee,dmt,said}. 
Correlations between the fitting
parameters were taken into account in the errors estimated by the
fitting routine \cite{stave_thesis,minuit}.  The
fits were performed using the spectrometer cross sections and  were the same (within the errors) whether or not the
$\sigma_{LT'}$ data were included.  In addition, the fits used the
entire $I=3/2$ amplitude so that the unitarity of each model was
preserved.  At resonance, these $I=3/2$ amplitudes are purely
imaginary due to the Fermi-Watson theorem \cite{drechsel_tiator}.  
The fits were performed at the same value of $W$ at which the data were taken. 
The models were then used to extrapolate the value of the
multipoles at $W=1232$ MeV.  
Following our
previous work\cite{cnp,sparveris} we took the final results to be the
average of these model determinations and estimated the model dependent
error in the resonance amplitudes by taking  the RMS deviation of the
values \cite{stave_thesis}. We believe that this is reasonable since
the chosen models represent state of the art calculations and also a
variety of different approaches.  The results of the fits for the
resonant multipoles along with the EFT predictions are presented in
Table~\ref{table1} along with the original values for several
models. We also present the average fitted values for the four
reaction models considered here. The differences between these values
represent the model dependence due to the different background
multipoles. 
The effect of background amplitudes on the resonant amplitudes
was studied and determined to have an effect approximately the
same size as the model to model RMS deviation.  This study is detailed in
Refs. \cite{stave_thesis} and \cite{stave_etal}.
For the result of this experiment we take  the average
values of the fitted multipoles using each  model along with both the
experimental and model error. It can be seen that the model and
experimental errors are approximately the same magnitude. There is
generally good agreement between the EFT predictions and our
experimental results. This also indicates the importance of the pion
contribution to these amplitudes.
It can also be seen from Table~\ref{table1} 
that the dispersion between the original
model calculations for the quadrupole amplitudes has been
considerably reduced by the fitting. For the EMR the RMS deviation in
the original models is  reduced from 0.56\% to 0.20\%. For the CMR the
RMS deviation in the original models is reduced even more, from
0.82\% to 0.26\%.

Figure~\ref{fig:Q06} shows the evolution of the multipole ratios at low
$Q^2$.  There is reasonable consistency of the results from the
different laboratories.  The plotted lattice QCD results, with a linear pion mass extrapolation\cite{alexandrou}, are in general agreement with the data for the EMR but disagree for the CMR by a wide margin. The EFT analysis of Pascalutsa and Vanderhaeghen(PV) \cite{pasc} indicates that a linear extrapolation is
close to the data for the EMR but results in a considerable
underestimation of the CMR.  The results of
the two chiral calculations \cite{gail_hemmert,pasc} are also presented
in Fig.~\ref{fig:Q06}. As these are effective
field theories they contain empirical low energy constants. For Gail and Hemmert this includes fits to  the
dominant $M_{1+}^{3/2}$ multipole for $Q^{2}\leq 0.2$ (GeV/c)$^2$ and for the EMR at the photon point ($Q^{2}=0$). In order to achieve the good overall agreement they had to employ one  higher  order  term with another empirical constant. As also can be seen from the large estimated errors of the Pascalutsa and Vanderhaeghen EFT calculation\cite{pasc} a treatment of the next higher order term is required.
In Fig.~\ref{fig:Q06} two
representative constituent quark models, the newer hypercentral quark model (HQM)\cite{hqm}, and an older non-relativistic calculation of Capstick
\cite{capstick_karl}, have been included (the relativistic calculations are in even worse agreement with experiment). These curves are representative of quark models which typically
under-predict the dominant $M_{1+}^{3/2}$ multipole by $\simeq$ 30\%
and underestimate the EMR and CMR by an order of magnitude, even
predicting the wrong sign. One solution to this problem has been to add pionic degrees of freedom to quark models 
\cite{quark_pion}. 
All of these models treat the $\Delta$ as a bound state and therefore do not have the $\pi$N continuum (i.e., no background amplitudes) so that cross sections are not calculated. The Sato-Lee \cite{sato_lee} and DMT \cite{dmt} dynamical reaction models with pion cloud effects bridge this gap    
and  are in qualitative agreement with the $Q^{2}$ evolution of the data. These models calculate the virtual
pion cloud contribution dynamically but have an empirical
parameterization of the inner (quark) core contribution which gives
them some flexibility in these observables.  By contrast the empirical
MAID \cite{maid1} and SAID \cite{said} represent fits to other data
with a smooth $Q^2$ dependence. 

One way to see the major role played by the pion cloud contribution to
the resonant multipoles is that for this case the expected scale for
the $Q^{2}$ evolution is $m_\pi^2=0.02$ GeV$^2$. In these units the
range of the present experiment from $Q^{2}$ from 0.060 to 0.20
GeV$^{2}$/$c^{2}$ is 3 to 10 units. Therefore it is not surprising that one
should see relatively large changes in the predicted  $Q^2$ evolution
of the resonant multipoles as is shown in Figure~\ref{fig:Q06}. It is
also clear that there is  significant model dependence in these
predictions.

In conclusion, the new data are at the lowest measured $Q^2$ for
modern electroproduction where the
dominant pionic contribution is predicted to be increasing. This  $Q^{2}$ region is sufficiently low to be able to test chiral effective calculations. These results are in qualitative agreement with lattice calculations with a chiral extrapolation to the physical pion mass \cite{pasc}, with recent chiral perturbation
theory calculations \cite{gail_hemmert,pasc} and with dynamical models which explicitly include
the pion cloud \cite{sato_lee,dmt}.  However, all of these calculations require refinements in order to obtain quantitative agreement with experiment. This includes lattice calculations with lighter pion masses and the next order in effective field theory.

We thank
 L. Tiator,  D. Drechsel, T.-S. H. Lee, V. Pascalutsa, M. Vanderhaeghen,
T. Gail and T. Hemmert for their assistance with valuable discussions and for sharing their unpublished work.  
This work is supported at Mainz by the Sonderforschungsbereich 443 of
 the Deutsche Forschungsgemeinschaft (DFG),  
University of Athens by the Program PYTHAGORAS of the Greek ministry of 
Education (co-funded by the European Social Fund and National Resources),
and at MIT by the  U.S. DOE under Grant No. DE-FG02-94ER40818.

\begin{table*}
\caption{\label{table1}Values of EMR(\%), CMR(\%), and $M_{1+}^{3/2}$ (in $10^{-3}/m_{\pi^+}$) extracted from these data with three resonant parameter fits using the 
 SAID \cite{said},
  MAID 2003 \cite{maid1}, 
  Sato-Lee(SL) \cite{sato_lee}, and 
  DMT \cite{dmt} 
  models at the $\Delta$ resonance, $W=1232$ MeV, at $Q^2=0.060$ (GeV/c)$^2$.  The
  original model predictions are in the square brackets.
  The first error is the statistical and
  cut errors added in
  quadrature. For $M_{1+}^{3/2}$ the second error is the systematic error (for the EMR and CMR they
are negligible).  For the average, the
  third error is the model error defined as the RMS deviation of the results
  from the four different models.
  The bottom two lines are the EFT predictions of Gail and Hemmert (GH)
  \cite{gail_hemmert} and  Pascalutsa and Vanderhaeghen (PV)
  \cite{pasc}.
}
\begin{center}
\begin{tabular}{cccc}
\hline\noalign{\smallskip}
   & EMR (\%) & CMR (\%) & $M_{1+}^{3/2}$ \\
\noalign{\smallskip}\hline\noalign{\smallskip}
SAID & $-2.18\pm0.31$[-1.80] & $-4.87 \pm 0.29$[-5.30]  & $40.81 \pm 0.29 \pm 0.57$[40.72]  \\
SL & $-2.26 \pm 0.30$[-2.98]  & $-4.46 \pm 0.25$[-3.48]  & $40.20 \pm 0.27 \pm 0.56$[41.28]  \\
DMT & $-2.11 \pm 0.28$[-2.84]  & $-4.85 \pm 0.26$[-5.74]  & $40.78 \pm 0.27 \pm 0.57$[40.81]  \\
MAID & $-2.56 \pm 0.27$[-2.16]  & $-5.07 \pm 0.26$[-6.51]  & $39.51 \pm 0.26 \pm 0.57$[40.53] \\  
\noalign{\smallskip}\hline\noalign{\smallskip}
Avg. & $-2.28 \pm 0.29_{\rm stat+sys} \pm 0.20_{\rm model}$  & 
$-4.81 \pm 0.27_{\rm stat+sys} \pm 0.26_{\rm model}$ & 
$40.33 \pm 0.27_{\rm stat} \pm 0.57_{\rm sys} \pm 0.61_{\rm model}$\\
\noalign{\smallskip}\hline\noalign{\smallskip}
GH   & -2.66 & -6.06 & 41.15 \\
PV   & $-2.88 \pm 0.70$ & $-5.85 \pm 1.40$ & $39.75 \pm 3.87$\\
\noalign{\smallskip}\hline
\end{tabular}
\end{center}
\end{table*}

\begin{figure*}
\begin{center}
\includegraphics[angle=0,height=6cm,width=18cm]{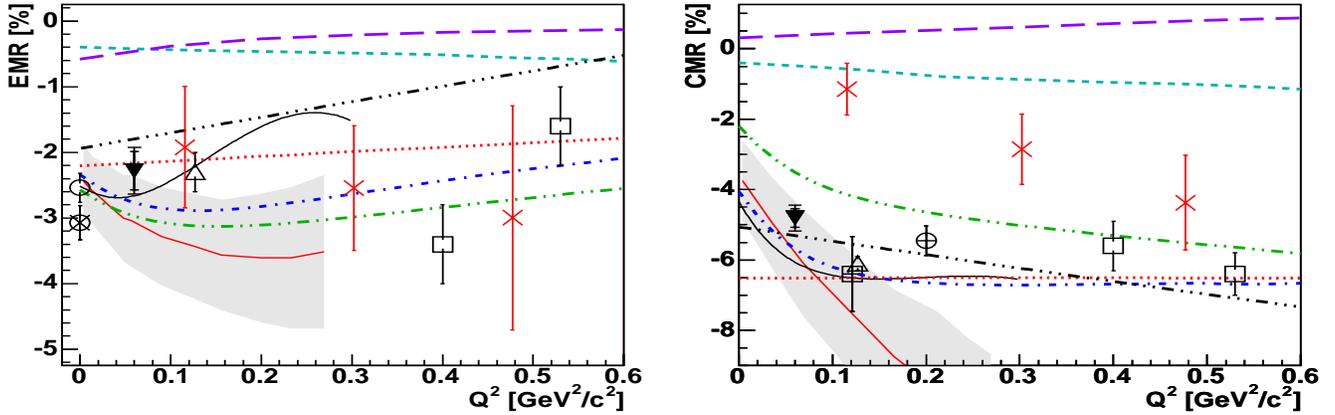}
\end{center}
\caption{\label{fig:Q06}(Color online)
The low $Q^2$ dependence of the EMR and CMR at $W=1232$ MeV for the $\gamma^{*} p \rightarrow \Delta$ reaction. The $\blacktriangledown$ symbols are our data
points and include the experimental and model errors (see
Table~\ref{table1}) 
added in quadrature. The other data are: the photon point
$\bigcirc$ \cite{beck} and $\otimes$ \cite{blanpied}, CLAS $\square$ \cite{joo},  Bates $\triangle$ \cite{sparveris}, Elsner $\bigoplus$\cite{elsner}, and Pospischil $\boxplus$ \cite{pospischil}. The lattice QCD calculations with linear pion mass extrapolations are shown as $\times$ \cite{alexandrou} and also the  recent chiral perturbation calculations of Pascalutsa and
  Vanderhaeghen (PV) (see EFT in Fig.~\ref{fig:XS:Sig0_LT}) \cite{pasc} 
and Gail and Hemmert (GH) (black solid line) \cite{gail_hemmert}.
The other curves represent the same models as in Fig.~\ref{fig:XS:Sig0_LT}. 
The HQM (long-dashed line) \cite{hqm} and
Capstick (short-dashed line) \cite{capstick_karl}  quark models have been included. }
\end{figure*}


\end{document}